# ST Chamaeleontis and BP Coronae Australis: Two Southern Dwarf Novae Confirmed as Z Cam Stars

**Mike Simonsen**
AAVSO, 49 Bay State Rd. Cambridge MA 02139 mikesimonsen@aavso.org

**Terry Bohlsen**
Mirranook Observatory, Booroolong Rd., Armidale, NSW, Australia;
terry.bohlsen@bigpond.com

**Franz-Josef Hambsch**
Oude Bleken 12, Mol, 2400, Belgium hambsch@telenet.be

**Rod Stubbings**
Tetoora Road Observatory, Tetoora Road, Vic, Australia; stubbo@sympac.com.au

**Abstract** Z Camelopardalis (Z Cam) stars are a subset of dwarf novae distinguished by the occurrence of "standstills" – periods of relative constant brightness one to one and a half magnitudes fainter than maximum brightness. As part of an ongoing observing campaign, the Z CamPaign, the authors focused attention on several Z Cam suspects in the southern hemisphere. Two stars, BP Coronae Australis and ST Chamaeleontis were found to exhibit standstill behavior in 2013, thus confirming them as Z Cam type systems. This adds two more bona fide members to the 19 confirmed Z Cams, bringing the total to 21.

1. Introduction
Dwarf novae are a subset of cataclysmic variables, close binary pairs consisting of a white dwarf (WD) primary and a main sequence secondary. The primary accretes matter from the secondary star through the inner Lagrangian point forming a disk around the WD primary in non-magnetic systems.

Z Cam stars are a subset of dwarf novae that, along with the typical dwarf novae outbursts and relatively brief periods of quiescence, occasionally display standstills in their light curves. It is these standstills that set the Z Cam stars apart from all other dwarf novae (Simonsen et al. 2014).

Since their discovery in the late 19$^{th}$ century, our understanding of dwarf novae has grown immeasurably. Along the way, our classifications and definitions have evolved to what they are today, causing some stars to be re-classified. This is especially true for the Z Cam stars whose definition has evolved considerably since being introduced over 80 years ago.

## 2. History

As with nearly all the bona fide and suspected Z Cams in the literature, the classifications of BP CrA and ST Cha have been the subject of confusion and debate for many years. This has been caused, in large part, by the evolution of the definition of Z Cams over time (Simonsen et al. 2014). In the case of many southern Z Cams and suspects there has historically been sparse coverage, which has led to misclassifications based on incomplete light curves and a lack of reliable information.

### 2.1. ST Cha

ST Cha was discovered in 1934 as an irregular variable (Luyten 1934). During the next 79 years it became variously classified as rapidly irregular, RW Aur, IS, T Tauri and a dwarf nova. Included in a study of RW Aurigae stars in 1974 (Glass and Penston 1974), ST Cha was not detected in the JHKL infrared bands. However, it was not included in the list of stars considered not to be RW Aur stars in this paper. Still considered a pre-main sequence star in 1975, ST Cha was classified as a T Tauri star (IBVS 1049). UBVRI photometry obtained in 1986 (Cieslinski, Steiner, Jablonski 1986) suggested a dwarf novae classification. Downes and Shara adopted this classification in their Catalog and Atlas of Cataclysmic Variables (Downes et al. 2001). Another noteworthy CV catalog (Ritter and Kolb 2003) has it ambiguously classified as a possible nova-like variable or suspected dwarf novae that might be a Z Cam sub-type. The General Catalogue of Variable Stars still lists ST Cha as an IS type, a rapidly irregular variable not associated with nebulosity, as of December 2013 (Samus et al. 2007).

### 2.2. BP CrA

BP CrA was initially classified as an irregular variable of the SS Cygni (UGSS) type from plates from the Union Observatory, Johannesburg, South Africa (van Gent 1933). In 1960 it was suspected of being a Z Cam type variable (Petit 1960), but in 1961 it was re-classified as a possible RW Aur type variable by the same author (Petit 1961). It was later observed in a systematic attempt to determine which southern cataclysmic variables (CVs) might be binary stars by monitoring for eclipses (Mumford 1971). BP CrA is listed as a U Geminorum star in that paper. No eclipses were found. BP CrA's specific sub-classification remained unclear in papers published even after the nature of CVs was better understood, with some listing it as simply U Gem (Echevarria and Jones 1983) and others classifying it as a member of the Z Cam type (Vogt and Bateson 1982, Bruch 1983, Pretorius, Downes et al. 2001, Warner and Woudt 2006). The General Catalog of Variable Stars still classified BP CrA as a suspected Z Cam as late as 2013 (Samus et al. 2007). To this day, the orbital period of BP CrA is unknown.

## 3. Standstills of ST Cha and BP CrA

The light curves in this paper are from *VSTAR* (Benn 2012). Black dots are visual data. Green dots are CCDV data.

ST Cha

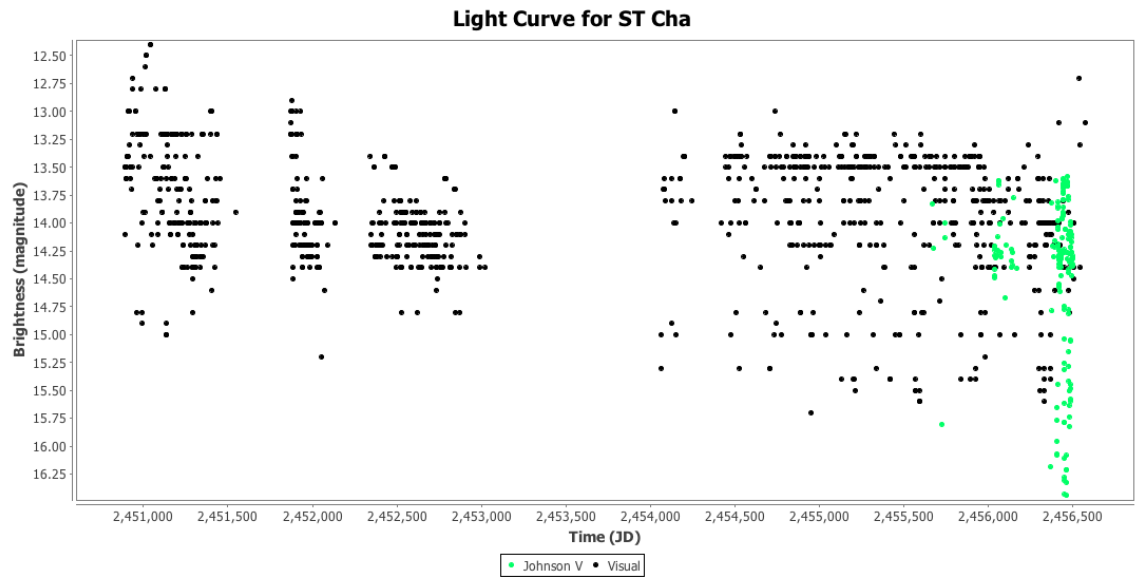

Figure 1. This is the long-term AAVSO light curve for ST Cha. It appears to show more or less common dwarf novae behavior from JD 2450892 to 2452134 (March 20, 1998–August 13, 2001). Then there is a period that could be interpreted as a standstill, or even nova-like behavior from JD to 2452353 to 2453023 (March 17, 2002- January 18, 2004). After a significant gap in the data, typical dwarf novae behavior, with outbursts and quiescences, resumes from JD 2454062 (November 22, 2006) to the present.

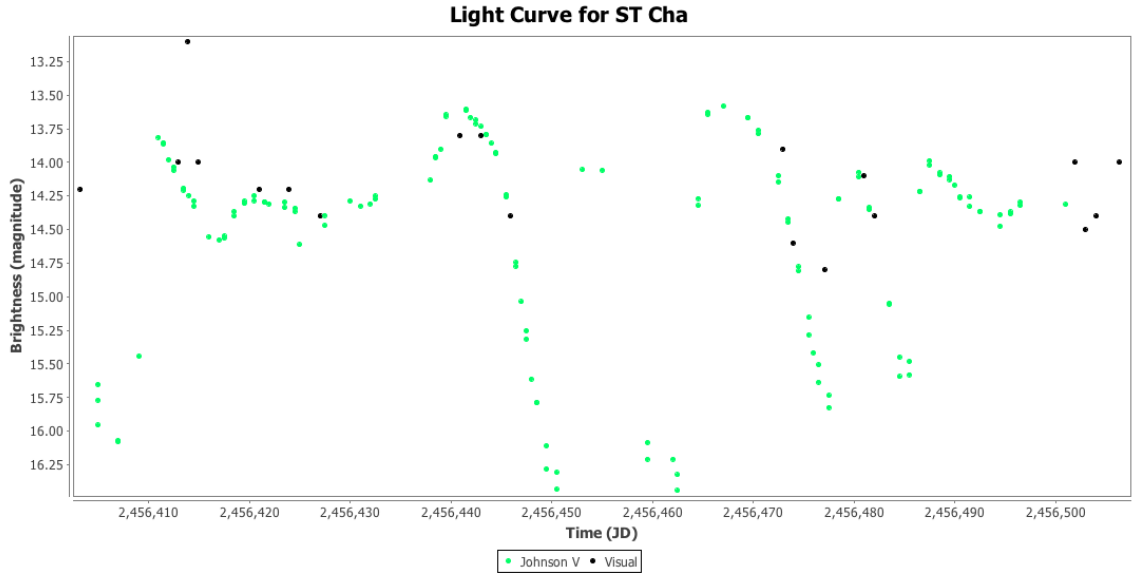

Figure 2. Embedded in the light curve from 2013, the almost daily snapshot CCDV observations reveal a complex behavior featuring standstills followed by outbursts and minima in rapid succession, reminiscent of V513 Cas and IW And (Simonsen 2011, Szkody et al. 2013), confirming the Z Cam classification of ST Cha.

BP CrA

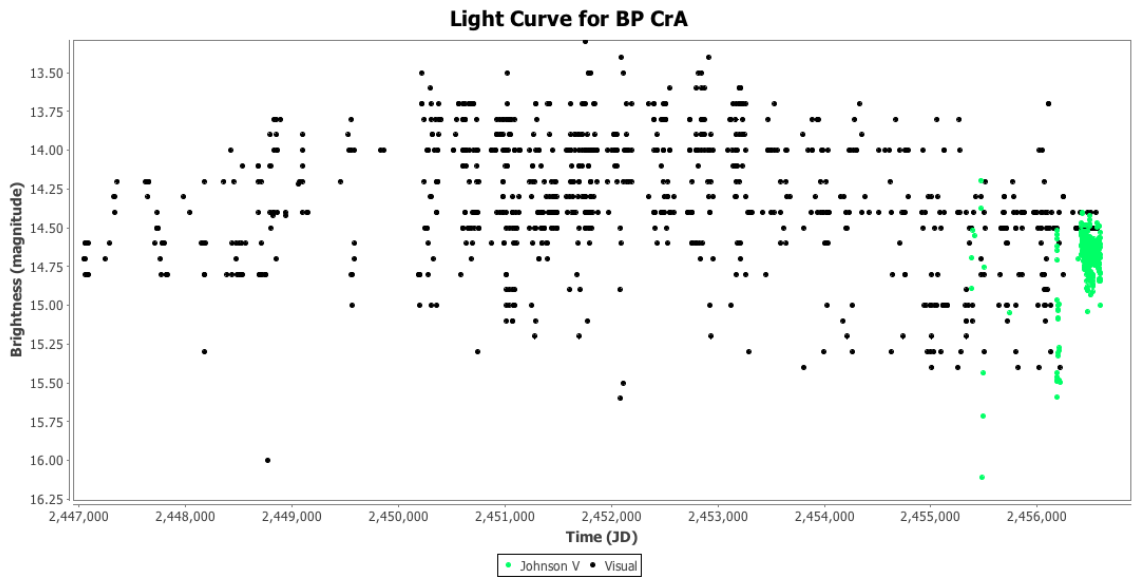

Figure 3. The long-term AAVSO light curve for BP CrA showing the spotty coverage dating back to 1987 and the recent concentrated effort beginning just before the first recorded standstill. There are no obvious standstills in the data until year 2013 (beginning JD 2456362).

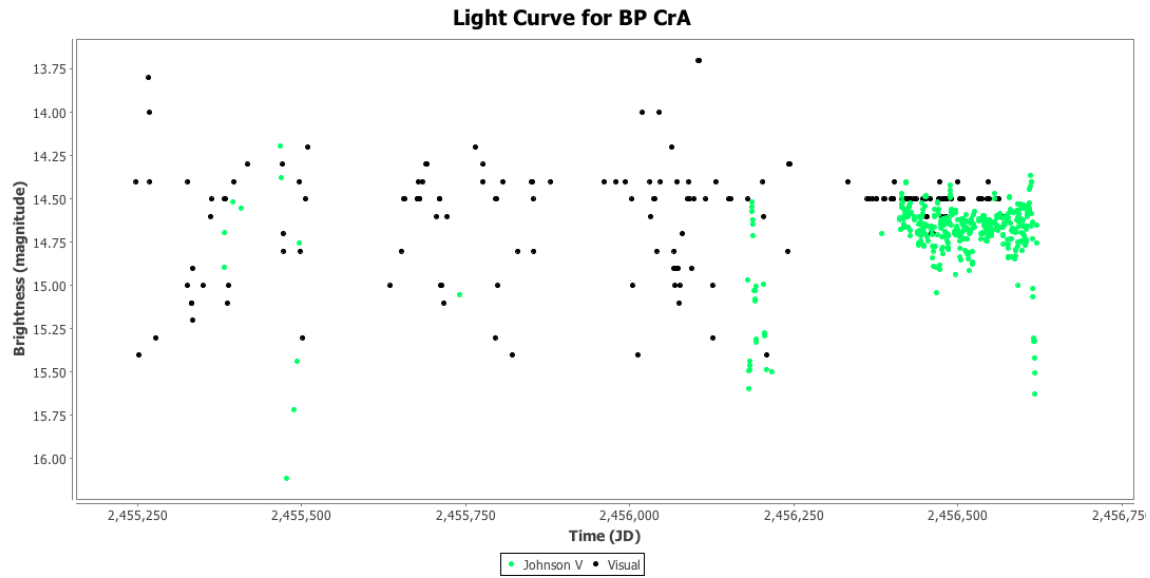

Figure 4. The detailed AAVSO light curve, showing the first recorded standstill of BP CrA beginning JD 2456362 and ending JD 2456615 (March 10, 2013 – November 18, 2013). This unambiguous example of a standstill confirms the status of BP CrA as a Z Cam dwarf nova.

## 4. Conclusions

Newly acquired data clearly demonstrates the existence of standstills in the light curves of ST Cha and BP CrA, making them bona fide members of the Z Cam class of dwarf novae. This classification and their other known properties are listed in Table 1.

## 5. Acknowledgements

We acknowledge with thanks the variable star observations from the AAVSO International Database contributed by observers worldwide and used in this research. This research has made use of NASA's Astrophysics Data System.

Table 1. Properties of two new Z Cam systems

| Name | RA (2000) | Dec (2000) | Max. | Min. | Type | Orbital period |
|---|---|---|---|---|---|---|
| ST Cha | 10 47 15.91 | -79 26 07.2 | 12.4V | 16.4V | UGZ | 6.84 hrs |
| BP CrA | 18 36 50.89 | -37 25 53.6 | 12.7V | 16.4V | UGZ | unknown |